\documentclass[12pt]{article}

\usepackage{amssymb,amsmath,amscd,amsthm}

\usepackage{graphicx}

\newtheorem{theorem}{Theorem}[section]

\newtheorem{lemma}[theorem]{Lemma}
\newtheorem{proposition}[theorem]{Proposition}
\newtheorem{definition}[theorem]{Definition}

\begin{document}

\title{An Inverse Problem for Gibbs Fields with Hard Core Potential}

\author{ Leonid Koralov\footnote{Partially supported by NSF Research Grant} \\[1pt]
\normalsize Department of Mathematics\\[-4pt] \normalsize
University of Maryland\\[-4pt] \normalsize College Park, MD
20742-4015\\[-4pt] \normalsize koralov@math.umd.edu\\[-4pt] }

\date{}
\maketitle

\begin{abstract}
It is well known that for a regular stable potential of pair
interaction and a small value of activity one can define the
corresponding Gibbs field (a measure on the space of
configurations of points in~$\mathbb{R}^d$).

In this paper we consider a  converse problem. Namely, we show
that for a sufficiently small constant $\overline{\rho}_1$ and a
sufficiently small function $\overline{\rho}_2(x)$, $x \in
\mathbb{R}^d$, that is equal to zero in a neighborhood of the
origin, there exist a hard core pair potential, and a value of
activity, such that $\overline{\rho}_1$ is the density and
$\overline{\rho}_2$ is the pair correlation function of the
corresponding Gibbs field.
\end{abstract}
\medskip
 {\bf Key words}: Gibbs Field, Gibbs Measure, Cluster Functions,
Pair Potential, Correlation Functions, Ursell Functions.

\medskip
{\bf MSC Classification}: 60G55, 60G60.

\section {Introduction}
\label{se1} Let us consider a translation invariant measure $\mu$
on the space of particle configurations on the space $
\mathbb{R}^d$. An $m$-point correlation function
$\rho_m(x_1,...,x_m)$ is the probability density for finding $m$
different particles at locations $x_1,...,x_m \in \mathbb{R}^d$.
The following natural question has been extensively discussed in
physical and mathematical literature: given $\rho_1(x_1) \equiv
\overline{\rho}_1$ and $\rho_2(x_1,x_2) = \overline{\rho}_2(x_1 -
x_2)$, does there exist a measure $\mu$, for which these are the
first correlation function (density) and the pair correlation
function, respectively?

In the series of papers \cite{L1}-\cite{L3} Lenard provided a set
of relations on the functions $\rho_m$ which are necessary and
sufficient for the existence of such a measure. However, given
$\rho_1$ and $\rho_2$, it is not clear how to check if there are
some $\rho_3, \rho_4, ...$ for which these relations hold.

There are several recent papers which demonstrate the existence of
particular types of point processes (measures on the space of
particle configurations), which correspond to given $\rho_1$ and
$\rho_2$ under certain conditions on $\rho_1$ and $\rho_2$. In
particular, one dimensional point processes of renewal type are
considered  by Costin and Lebowitz in \cite{CL}, while
determinantal processes are considered by Soshnikov in \cite{So}.
In \cite{AS} Ambartzumian and Sukiasian prove the existence of a
point process corresponding to a sufficiently small density and
correlation function. Recently Costin and Lebowitz \cite{CL}, and
Caglioti, Kuna, Lebowitz, and Speer \cite{CK} provided various
generalizations of their results. In \cite{ST} Stillinger and
Torquato consider fields over a space with finitely many points.
Besides, for the lattice model, Stillinger and Torquato discuss
possible existence of a pair potential for a given density and
correlation function using cluster expansion without addressing
the issue of convergence.

In this paper we show that if $\rho_1$ and $\rho_2$ are small (in
a certain sense), and $\rho_2$ is zero in a neighborhood of the
origin, there exists a measure on the space of configurations for
which $\rho_1$ is the density and $\rho_2$ is the pair correlation
function. Moreover, this measure is the Gibbs measure
corresponding to some hard core pair potential and some value of
activity. In a sense, this is the converse of the classical
statement that a given potential of pair interaction and a
sufficiently small value of activity determine a translation
invariant Gibbs measure on the space of particle configurations in
$ \mathbb{R}^d$ and the sequence of infinite volume correlation
functions.

In our earlier paper \cite{K} we obtained a similar result for
lattice systems. In this paper we shall demonstrate that we can
extend those results to particle systems in $\mathbb{R}^d$ if we
assume that $\overline{\rho}_2$ is equal to zero in a neighborhood
of the origin.

\section{Notations and Formulation of the Result}
\label{se2}

Let $ \Phi(x)$, $x \in \mathbb{R}^d$ be a hard core potential of
pair interaction, that is $\Phi(x)$ is a measurable real-valued
function for $|x| \geq R$, and $\Phi(x) = +\infty$ if $|x| < R$,
where $R > 0$. Without loss of generality we may put $R = 1$. We
assume that $\Phi(x) = \Phi(-x)$ for all $x$. Let $U(x_1,...,x_n)
= \sum_{1 \leq i < j \leq n} \Phi(x_i - x_j)$ be the total
potential energy of the configuration $(x_1,...,x_n)$.

We call two pair potentials equivalent if they are equal almost
everywhere. When we say that an inequality involving $\Phi$ holds,
it will mean that the inequality is true for some representative
of the equivalence class.

 A potential of pair interaction is said to be stable if there is
 a constant $c$ (and an element of the equivalence class of $\Phi$) such that
\[
U(x_1,...,x_n) \geq - n c~~~{\rm for}~{\rm all}~n~~{\rm
and}~~x_1,...,x_n \in \mathbb{R}^d.
\]

A potential of pair interaction is said to be regular if it is
essentially bounded from below and satisfies
\[
\int_{\mathbb{R}^d} |\exp(- \Phi(x)) -1 | d x < +\infty.
\]

When discussing the properties of $\Phi$, it will be convenient to
consider the following two classes of sets. We shall say that a
set $\mathcal{X}$ belongs to the class $ \mathcal{C}$ if it
consists of a finite number of points in $\mathbb{R}^d$, and has
the property that $|x-y| \geq 1$ if $x,y \in \mathcal{X}, x \neq
y$. We shall say that $\mathcal{X}$ belongs to the class $
\mathcal{C}_0$ if it consists of a finite number of points in
$\mathbb{R}^d \setminus B_1$, where $B_1$ is the open unit ball
centered at the origin, and has the property that $|x-y| \geq 1$
if $x,y \in \mathcal{X}, x \neq y$.

For a measurable function $f$ defined on $\mathbb{R}^d$ we have
the estimates
\begin{equation} \label{inte1}
\int_{\mathbb{R}^d} |f(x)| d x \leq {\rm Vol}(B_1) \sup_{
\mathcal{X} \in \mathcal{C}} \sum_{x \in \mathcal{X}}
|f(x)|,~~\int_{\mathbb{R}^d \setminus B_1 } |f(x)| d x \leq {\rm
Vol}(B_1) \sup_{ \mathcal{X} \in \mathcal{C}_0} \sum_{x \in
\mathcal{X}} |f(x)|.
\end{equation}
Let us prove the first estimate. If $|f|$ is unbounded then the
right-hand side is equal to~$+\infty$, and the estimate follows.
Assume that $|f|$ is bounded. Let $R, \varepsilon > 0$, and $B_R$
be the ball of radius $R$ centered at the origin. Let $x_1$ be a
point in $B_R$ such that $|f(x_1)| \geq \sup_{x \in B_R} |f(x)| -
\varepsilon/2$. Let $B(x_1)$ be the unit ball centered at $x_1$.
Now we can inductively define points ${x_2,...,x_n \in B_R}$ as
follows. Let $x_k$, $2 \leq k \leq n$, be a point in $B_R
\setminus (B(x_1) \cup ...\cup B(x_{k-1}))$ such that
\[
|f(x_k)| \geq \sup_{x \in B_R \setminus (B(x_1) \cup ...\cup
B(x_{k-1}))} |f(x)| - \varepsilon/2^k.
\]
Here $n$ is such that the union of the unit balls
$B(x_1),...,B(x_n)$ covers $B_R$. Then the integral of $|f|$ over
$B_R$ is estimated by ${\rm Vol}(B_1) ((|f(x_1)|+...+|f(x_n)| +
\varepsilon)$. Since $R$ and $\varepsilon$ were arbitrary, this
implies the first estimate in (\ref{inte1}). The second estimate
is completely similar.

 Let $ \Lambda $ be a finite subset of $ \mathbb{R}^d$. The grand
canonical ensemble is defined by a measure on $ \bigcup_{n =
0}^\infty \Lambda^n$, whose restriction on $\Lambda^n$ has the
density
\[
\nu (x_1,...,x_n) = \frac{z^n}{n !} e^{-U(x_1,...,x_n)}~.
\]
The parameter $z >0$ is called the activity. The inverse
temperature, which is the factor usually present in front of the
function $U$, is set to be equal to one (or, equivalently,
incorporated into the function $U$). The total mass of the measure
is the grand partition function
\[
\Xi(\Lambda, z, \Phi) = \sum_{n = 0}^\infty \frac{z^n}{n !}
\int_{(x_1,...,x_n) \in \Lambda^n}  e^{-U(x_1,...,x_n)}~.
\]
The $m$-point correlation function is defined as the probability
density for finding $m$ different particles at positions
$x_1,...,x_m \in \Lambda$,
\[
\rho^{\Lambda}_m(x_1,...,x_m)  = \Xi(\Lambda, z, \Phi)^{-1}
\sum_{n = 0}^\infty \frac{z^{m+n}}{n !} \int_{(y_1,...,y_n) \in
\Lambda^n} e^{-U(x_1,...,x_m,y_1,...,y_n)}~.
\]
The corresponding measure on the space of all configurations of
particles on the set $\Lambda$ (Gibbs measure) will be denoted by
$\mu^\Lambda$. Given another set $\Lambda_0 \subseteq \Lambda$, we
can consider the measure $\mu^{\Lambda}_{\Lambda_0}$ obtained as a
restriction of the measure $\mu^\Lambda$ to the set of particle
configurations on $\Lambda_0$.

 Given a potential of pair interaction $\Phi(x)$, we define $g(x)
= e^{-\Phi(x)} - 1$, $x \in \mathbb{R}^d$. We shall make the
following three assumptions:
\begin{equation} \label{assum1}
g(x) = g(-x) \geq -a > -1~~~{\rm almost}~{\rm surely}~~~ {\rm
for}~~ |x| \geq 1.
\end{equation}
\begin{equation}
\label{assum2}
 g(x) = -1~~~{\rm almost}~{\rm surely}~~~{\rm for}~~|x| < 1.
\end{equation}
For some element of the equivalence class of $g$ we have
\begin{equation} \label{assum3}
 \sup_{ \mathcal{X} \in \mathcal{C}_0} \sum_{x \in \mathcal{X}} |g(x)|
\leq c < +\infty.
\end{equation}

Clearly, any function $g(x)$ which satisfies
(\ref{assum1})-(\ref{assum2}) defines a hard core potential of
pair interaction via
\[
\Phi(x) = - \ln(g(x) +1)~.
\]
From (\ref{assum3}) it easily follows that $\Phi$ is regular and
stable. Indeed, it is regular due to (\ref{inte1}), while the
stability follows from
\[
U(x_1,...,x_n) = -\frac{1}{2} \sum_{i = 1}^n \sum_{j \neq i} \ln
(1+ g(x_j  -x_i)) \geq
\]
\[
 -\frac{1}{2} n \sup_{ \mathcal{X} \in
\mathcal{C}_0} \sum_{x \in \mathcal{X}} |g(x)| \geq -\frac{1}{2} n
c~~~~{\rm if}~|x_i - x_j| \geq 1~~~{\rm for}~~i \neq j,
\]
and
\[
U(x_1,...,x_n) = + \infty ~~~~{\rm if}~|x_i - x_j| < 1~~~{\rm
for}~~i \neq j.
\]

 It is well known (\cite{A}, \cite{B}) that for stable regular
pair potentials the following two limits exist  for sufficiently
small $z$ when $\Lambda \rightarrow \mathbb{R}^d$ in a suitable
manner (for example, $\Lambda = [-k, k]^d$ and $k \rightarrow
\infty$):
\\ (a) There is a probability  measure $\mu^{\mathbb{R}^d}$ on the
space of all locally finite configurations on~$\mathbb{R}^d$, such
that
\begin{equation} \label{limit1}
\mu^\Lambda_{\Lambda_0} \rightarrow
\mu^{\mathbb{R}^d}_{\Lambda_0}~~~{\rm as}~~ \Lambda \rightarrow
\mathbb{R}^d
\end{equation}
for any finite set $\Lambda_0 \subset \mathbb{R}^d$.\\
 (b) All the correlation functions converge to
the infinite volume correlation functions. Namely,
\begin{equation} \label{limit2}
{\rm ess~sup}_{x_1,...x_m \in \Lambda_0} |
\rho^{\Lambda}_m(x_1,...,x_m) -\rho_m(x_1,...,x_m)| \rightarrow 0
~~~{\rm as} ~~ \Lambda \rightarrow \mathbb{Z}^d
\end{equation}
for any finite set $\Lambda_0 \subset \mathbb{R}^d$. The infinite
volume correlation functions $\rho_m$  are the probability
densities for finding $m$ different particles at positions
$x_1,...,x_m \in \mathbb{R}^d$ corresponding to the measure
$\mu^{\mathbb{R}^d}$.

To make these statements precise we formulate them as the
following lemma (here we take into account that
(\ref{assum1})-(\ref{assum3}) imply that the potential is regular
and stable).
\begin{lemma} (\cite{A}, \cite{B})
Assuming that (\ref{assum1})-(\ref{assum3}) hold, there is a
positive $ \overline{z} = \overline{z}(a,c)$, such that
(\ref{limit1}) and (\ref{limit2}) hold for all $0 < z \leq
 \overline{z}$ when
$\Lambda = [-k, k]^d$ and $k \rightarrow \infty$.
\end{lemma}
Thus,   a pair potential defines a sequence of infinite volume
correlation functions for sufficiently small values of activity.
Note that  all the correlation functions are translation
invariant,
\[
\rho_m(x_1,...,x_m) = \rho_m(x_1+a,...,x_m+a)~~~{\rm for}~{\rm
any}~a \in \mathbb{R}^d.
\]
Thus, $\rho_1$ is a constant, $\rho_2$ can be considered as a
function of one variable, etc. Let $\overline{\rho}_m$ be the
function of $m-1$ variables, such that
\begin{equation} \label{sme}
\rho_m(x_1,...,x_m) = \overline{\rho}_m (x_2 - x_1,...,x_m -x_1)~.
\end{equation}

The main result of this paper is the following theorem.
\begin{theorem} \label{main}
Let $0 <  r < 1$ be a constant. Let $\overline{\rho}_1$ be a
constant, and let $\overline{\rho}_2(x)$, $x \in \mathbb{R}^d$, be
a function that satisfies
\[
\overline{\rho}_2(x) = 0~~~{\rm for}~ |x| < 1;~~
\overline{\rho}_2(x) = \overline{\rho}_2(-x),~~{\it and}
\]
\[
\sup_{ \mathcal{X} \in  \mathcal{C}_0} \sum_{x \in \mathcal{X}}
|\overline{\rho}_2(x) - \overline{\rho}_1^2| \leq r
\overline{\rho}_1^2.
\]
Then for all sufficiently small values of $\overline{\rho}_1$
 there are a
potential $\Phi(x)$, which satisfies
(\ref{assum1})-(\ref{assum3}), and a value of activity $z$, such
that $\overline{\rho}_1$ and $\overline{\rho}_2(x)$ are the first
and the second correlation functions, respectively, for the system
defined by $(z, \Phi)$.
\end{theorem}

\noindent {\bf Remark.} As will seen from the proof of the
theorem, the pair potential and the activity corresponding to
given $\overline{\rho}_1$ and $\overline{\rho}_2(x)$ are unique,
if we restrict consideration to sufficiently small values of
$\Phi$ and $z$.
\\

 The outline of the proof is the following. In Sections
\ref{clust} and \ref{erp}, assuming that a pair potential and a
value of the activity exist, we express the correlation functions
(or, rather, the cluster functions, which are closely related to
the correlation functions) in terms of the pair potential and the
activity. This relationship can be viewed as an equation for
unknown $\Phi$ and $z$. In Section \ref{mainr} we use the
contracting mapping principle to demonstrate that this equation
has a solution. In Section \ref{urs} we provide the technical
estimates needed to prove that the right hand side of the equation
on $\Phi$ and $z$ is indeed a contraction.

\section{Cluster Functions and  Ursell Functions}
\label{clust}

In this section we shall obtain a useful expression for cluster
functions in terms of the pair potential. The cluster functions
are closely related to the correlation functions.
 Some of the general well-known facts will be
stated in this section without proofs. The reader is referred to
Chapter 4 of \cite{A} for a more detailed exposition.

Let $A$ be the complex vector space of sequences $\psi$,
\[
\psi = (\psi_m(x_1,...,x_m))_{m \geq 0}
\]
such that, for each $m \geq 1$, $\psi_m$ is an essentially bounded
measurable function on $\mathbb{R}^{md}$, and $\psi_0$ is a
complex number. It will be convenient to represent a finite
sequence $(x_1,...,x_m)$ by a single letter $X = (x_1,...,x_m)$.
We shall write
\[
\psi(X)  = \psi_m(x_1,...,x_m).
\]
Let now $\psi^1, \psi^2 \in A$. We define
\[
\psi^1 \ast \psi^2 (X) = \sum_{Y \subseteq X} \psi^1(Y) \psi^2(X
\backslash Y),
\]
where the summation is over all subsequences $Y$ of $X$ and $X
\backslash Y$ is the subsequence of $X$ obtained by striking out
the elements of $Y$ in $X$.
\\

\noindent {\bf Remark on Notation.}  Let us stress that the
inclusion $Y \subseteq X$ means here that $Y$ is a subsequence of
$X$, rather than a simple set-theoretic inclusion. Below we shall
also use the notation $X \cup Y$ for the sequence obtained by
adjoining $X$ and $Y$.
\\

Let $A_+$ be the subspace of $A$ formed by the elements $\psi$
such that $\psi_0 = 0$. Let $ \mathbf{1}$ be the unit element of
$A$ ( $ \mathbf{1}_0 = 1, \mathbf{1}_m \equiv 0$ for $m \geq 1$).

We define the mapping $\Gamma$ of $A_+$ onto $\mathbf{1} + A_+$ :
\[
\Gamma \varphi = \mathbf{1} + \varphi + \frac{\varphi \ast
\varphi}{2!} + \frac{\varphi \ast \varphi \ast \varphi}{3!} + ...
\]
The mapping $\Gamma$ has an inverse $\Gamma^{-1}$ on $\mathbf{1} +
A_+$:
\[
\Gamma^{-1}(\mathbf{1} + \varphi') = \varphi' - \frac{\varphi'
\ast \varphi'}{2} + \frac{\varphi' \ast \varphi' \ast \varphi'}{3}
- ...
\]
It is easy to see that $\Gamma \varphi (X)$ is the sum of the
products $\varphi(X_1)...\varphi(X_r)$ corresponding to all the
partitions of $X$ into subsequences $X_1,...,X_r$. If $\varphi \in
A_+$ and $\psi = \Gamma \varphi$, the first few components of
$\psi$ are
\[
\psi_0 = 1;~~~~\psi_1(x_1) = \varphi_1(x_1);~~~~\psi_2(x_1, x_2) =
\varphi_2(x_1, x_2) + \varphi_1(x_1) \varphi_1(x_2).
\]
Let $\Phi$ be a pair correlation function which satisfies
(\ref{assum1})-(\ref{assum3}), and let $z \leq \overline{z}(a,c)$.
Note that the sequence of correlation functions $\rho =
({\rho_m})_{m \geq 0}$ (with $\rho_0 = 1$) is an element of
$\mathbf{1} + A_+$.
\begin{definition}
The cluster functions $\omega_m(x_1,...,x_m)$, $m \geq 1$ are
defined by
\[
\omega = \Gamma^{-1} \rho.
\]
\end{definition}
Thus,
\[
\omega_1(x_1) = \rho_1(x_1);~~~~\omega_2(x_1,x_2) =
\rho_2(x_1,x_2) - \rho_1(x_1) \rho_1(x_2),
\]
or, equivalently,
\[
\overline{\omega}_1 = \overline{\rho}_1;~~~~\overline{\omega}_2(x)
= \overline{\rho}_2(x) - \overline{\rho}_1^2~
\]
where $\overline{\omega}_m$ are defined by
\[
\omega_m(x_1,...,x_m) = \overline{\omega}_m (x_2 - x_1,...,x_m
-x_1)~.
\]

Let $\psi \in \mathbf{1} + A_+$ be defined by
\[
\psi_0 = 1;~~~~\psi_m(x_1,...,x_m) = e^{-U(x_1,...,x_m)}~.
\]
Define also
\[
\varphi = \Gamma^{-1} \psi~.
\]
\begin{definition}
The functions $\psi_m$ and $\varphi_m$ are called Boltzmann
factors and Ursell functions, respectively.
\end{definition}
\begin{lemma}
\label{expansion} (\cite{A}) There is a positive constant
$\overline{\overline{z}}(a,c)$ such that for all $z \leq
\overline{\overline{z}}(a,c)$ the cluster functions can be
expressed in terms of the Ursell functions as follows
\[
\omega_m(x_1,...,x_m) = z^m \sum_{n = 0}^{\infty} \frac{z^n}{n!}
\int_{(y_1,...,y_n) \in \mathbb{R}^{nd}}
\varphi_{m+n}(x_1,...,x_m,y_1,...,y_n)~.
\]
\end{lemma}
We shall later need certain estimates on the Ursell functions in
terms of the potential. To this end we obtain a recurrence formula
on a set of functions related to the Ursell functions. Given $X =
(x_1,...,x_m)$, we define the operator $D_X: A \rightarrow A$ by
\[
(D_X \psi)_n (y_1,...,y_n) = \psi_{m+n}(x_1,...,x_m,y_1,...,y_n)~.
\]
Then define
\[
\widetilde{\varphi}_X = \psi^{-1} \ast D_X \psi~,
\]
where $\psi$ is the sequence of Boltzmann factors, and $\psi^{-1}$
is such that $\psi^{-1} \ast \psi = \mathbf{1}$. It can be seen
that
\begin{equation}
\label{rel1} \varphi_{1+n}(x_1,y_1,...,y_n) =
\widetilde{\varphi}_{x_1}(y_1,...,y_n)
\end{equation}
and that the functions $ \widetilde{\varphi}_X $ satisfy a certain
recurrence relation, which we state here as a lemma.
\begin{lemma} (\cite{A}) \label{recurr}
Let $X  = (x_1,...,x_m)$, $Y = (y_1,...,y_n)$, and $Z =
(z_1,...,z_s)$ be a generic subsequence of $Y$, whose length will
be denoted by $|Z|$. The functions $ \widetilde{\varphi}_X $
satisfy the following recurrence relation
\begin{equation} \label{rec}
\widetilde{\varphi}_X (Y) = \exp(- \sum_{i = 2}^m \Phi(x_i- x_1))
\sum_{Z \subseteq Y} \prod_{j = 1}^{|Z|}( \exp(-\Phi(z_j -x_1))
-1) \widetilde{\varphi}_{Z \cup X \backslash x_1} (Y \backslash
Z)~,
\end{equation}
where $m \geq 1$, $n \geq 0$, and $\widetilde{\varphi}_X (Y) =
\mathbf{1}$ if $m = 0$.
\end{lemma}

\section{Equations Relating the Potential, the Activity, and the Cluster Functions}
\label{erp} In this section we shall recast the main theorem in
terms of the cluster functions and examine a system of equations,
which relates the first two cluster functions with the pair
potential and the activity.

First, Theorem \ref{main} can clearly be re-formulated as follows
\begin{proposition} \label{main2}
Let $0 < r < 1$ be a constant. Given any sufficiently small
constant $\overline{\omega}_1$ and any function
$\overline{\omega}_2(x)$, such that
\[
\overline{\omega}_2(x) =-\overline{\omega}_1^2~~~{\rm for}~ |x| <
1;~~ \overline{\omega}_2(x) = \overline{\omega}_2(-x),~~{\it and}
\]
\[
\sup_{ \mathcal{X} \in \mathcal{C}_0} \sum_{x \in \mathcal{X}}
|\overline{\omega}_2(x)| \leq r \overline{\omega}_1^2,
\]
there are a potential $\Phi(x)$,  which satisfies
(\ref{assum1})-(\ref{assum3}), and a value of activity $z$, such
that $\overline{\omega}_1$ and $\overline{\omega}_2(x)$ are the
first and the second cluster functions respectively for the system
defined by $(z, \Phi)$.
\end{proposition}

Consider the power expansions for $\omega_1$ and $\omega_2$, which
are provided by Lemma \ref{expansion}. Let us single out the first
term in both expansions. Note the translation invariance of the
functions $\omega_m$ and $\varphi_m$ and the fact that
$\varphi(x_1, x_2) = g(x_1 - x_2)$.
\begin{equation} \label{clust1}
\overline{\omega}_1 = z + z^2 \sum_{n=1}^\infty \frac{z^{n-1}}{n!}
\int_{(y_1,...,y_n) \in \mathbb{R}^{nd}}
\overline{\varphi}_{1+n}(y_1,...,y_n)~,
\end{equation}
\begin{equation} \label{clust2}
\overline{\omega}_2(x) = z^2 g(x) + z^3 \sum_{n=1}^\infty
\frac{z^{n-1}}{n!} \int_{(y_1,...,y_n) \in \mathbb{R}^{nd}}
\overline{\varphi}_{2+n}(x,y_1,...,y_n)~,
\end{equation}
where  $\overline{\varphi}_m$ is the function of $m-1$ variables,
such that
\[
\varphi_m(x_1,...,x_m) = \overline{\varphi}_m (x_2 - x_1,...,x_m
-x_1)~.
\]
 Let
\[
A(z,g) = \sum_{n=1}^\infty \frac{z^{n-1}}{n!} \int_{(y_1,...,y_n)
\in \mathbb{R}^{nd}} \overline{\varphi}_{1+n}(y_1,...,y_n)~,
\]
\[
B(z,g)(x) = \sum_{n=1}^\infty \frac{z^{n-1}}{n!}
\int_{(y_1,...,y_n) \in \mathbb{R}^{nd}}
\overline{\varphi}_{2+n}(x,y_1,...,y_n)~.
\]
Thus equations (\ref{clust1}) and (\ref{clust2}) can be rewritten
as follows
\begin{equation} \label{eqn1}
z = \overline{\omega}_1 - z^2 A(z,g)~,
\end{equation}
\begin{equation} \label{eqn2}
g = \frac{\overline{\omega}_2}{z^2} - z B(z,g)~.
\end{equation}
Instead of looking at (\ref{eqn1})-(\ref{eqn2}) as a formula
defining $\overline{\omega}_1$ and $\overline{\omega}_2$ by a
given pair potential and the activity, we can instead consider the
functions $\overline{\omega}_1$ and $\overline{\omega}_2$ fixed,
and $g$ and $z$ unknown. Thus, Proposition \ref{main2} follows
from the following.
\begin{proposition} \label{main3}
If $\overline{\omega}_1$ and $\overline{\omega}_2$ satisfy the
assumptions of Proposition \ref{main2}, then the system
(\ref{eqn1})-(\ref{eqn2}) has a solution $(z,g)$, such that the
function $g$ satisfies (\ref{assum1})-(\ref{assum3}) and $z \leq
\overline{z}(a,c)$.
\end{proposition}
\section{Proof of the Main Result}
\label{mainr} This section is devoted to the proof of Proposition
\ref{main3}. We shall need the following notations. Let $
\mathcal{G}$ be the space of measurable functions $g$, which
satisfy (\ref{assum1})-(\ref{assum3}) with some~${a,c < \infty}$.
Let
\begin{equation} \label{norm3}
||g|| =  \sup_{ \mathcal{X} \in \mathcal{C}_0} \sum_{x \in
\mathcal{X}} |g(x)|.
\end{equation}
(To be more precise, the space $ \mathcal{G}$ consists of
equivalence classes - we do not distinguish between functions
which are equal almost surely. We assume that the element of the
equivalence class that minimizes the expression in the right hand
side of (\ref{norm3}) is used in the definition of $||g||$).

Note that $||g||$ is not a norm, since $ \mathcal{G}$ is not a
linear space, however ${d(g_1,g_2) = ||g_1-g_2||}$ is a metric on
the space $ \mathcal{G}$. Let $ \mathcal{G}_c$ be the set of
elements of $ \mathcal{G}$ for which $||g|| \leq c$. Note that if
$c <1$ then all elements of $ \mathcal{G}_c$ satisfy
(\ref{assum1}) with $a =c$.

We also define $I^{a_1, a_2}_{z_0} = [a_1 z_0, a_2 z_0]$. Let $D =
I^{a_1, a_2}_{z_0} \times  \mathcal{G}_c$. Note that if $c < 1$
then $(z,g) \in D$ implies that $z \leq \min( \overline{z}(c,c) ,
\overline{\overline{z}}(c,c)) $ if $z_0$ is sufficiently small.
Thus, the infinite volume correlation functions and cluster
functions are correctly defined for $(z,g) \in D$ if $z_0$ is
sufficiently small.

 Let us define
an operator $Q$ on the space of pairs $(z,g) \in D$ by $Q(z,g) =
(z', g')$, where
\begin{equation} \label{eqn1a}
z' = \overline{\omega}_1 - z^2 A(z,g)~,
\end{equation}
\begin{equation} \label{eqn2a}
g'(x) = \frac{\overline{\omega}_2(x)}{z^2} - z B(z,g)(x)~~~{\rm
for}~~x \geq 1;~~~~g'(x)= -1~~~{\rm for}~~x < 1.
\end{equation}
We shall prove the following lemma.
\begin{lemma} \label{l1}
Let $0 < r < 1$ be a constant. There exist positive constants $a_1
< 1$, $a_2 >1$, and $c < 1$ such that the equation $(z,g) =
Q(z,g)$ has a solution $(z,g) \in D$  for all sufficiently small
$z_0$ if
\[
\overline{\omega}_1 = z_0;~~~ \overline{\omega}_2(x)
=-\overline{\omega}_1^2~~~{\rm for}~ |x| < 1;~~
\overline{\omega}_2(x) = \overline{\omega}_2(-x),~~{\it and}
\]
\[
\sup_{ \mathcal{X} \in \mathcal{C}_0} \sum_{x \in \mathcal{X}}
|\overline{\omega}_2(x)| \leq r \overline{\omega}_1^2.
\]
\end{lemma}

Before we prove this lemma, let us verify that it implies
Proposition \ref{main3}. Let $0 < r < 1$ be fixed, and let
 $\overline{\omega}_1$ and
$\overline{\omega}_2$ verify the assumptions of Lemma \ref{l1}.
Let $(z,g)$ be the solution of $(z,g) = Q(z,g)$, whose existence
is guaranteed by Lemma \ref{l1}. Let $\overline{\omega}'_1$ and
$\overline{\omega}'_2$ be the first two cluster functions
corresponding to the pair $(z,g)$. Note that $\overline{\omega}_1$
and $\overline{\omega}'_1$ satisfy the same equation
\[
z = \overline{\omega}_1 - z^2 A(z,g)~;~~~~
 z = \overline{\omega}'_1
- z^2 A(z,g)~.
\]
Therefore, $\overline{\omega}_1 = \overline{\omega}'_1$. The
functions  $\overline{\omega}_2$ and $\overline{\omega}'_2$ also
satisfy the same equation
\[
g(x) = \frac{\overline{\omega}_2(x)}{z^2} - z B(z,g)(x)~;~~~~ g(x)
= \frac{\overline{\omega}'_2(x)}{z^2} - z B(z,g)(x)~;~~~~ {\rm
for}~~x \geq 1.
\]
Thus, $\overline{\omega}_2(x) = \overline{\omega}'_2(x)$ for $x
\geq 1$. The fact that $\overline{\omega}_2(x) =
\overline{\omega}'_2(x)$ for $x < 1$  follows from
\[
\overline{\omega}_2(x) = - \overline{\omega}_1^2 =   -
{\overline{\omega}'_1}^2 = \overline{\omega}'_2(x)~~~{\rm for}~~x
< 1.
\]
Thus it remains to prove Lemma \ref{l1}. The proof will be based
on the fact that for small $z_0$ the operator $Q: D \rightarrow D$
is a contraction in an appropriate metric.  Define
\[
d_{z_0}(z_1,z_2) = \frac{h |z_1 -z_2|}{z_0}~.
\]
The value of the constant $h$ will be specified later. Now the
metric on $D$ is given by
\[
\rho((z_1,g_1),(z_2,g_2)) =  d_{z_0}(z_1,z_2) + d(g_1,g_2)~.
\]
Lemma \ref{l1} clearly follows from the contracting mapping
principle and the following lemma
\begin{lemma} \label{l2}
Let $0 < r < 1$ be a constant. There exist positive constants $a_1
< 1$, $a_2 >1$, and $c < 1$ such that for all sufficiently small
$z_0$ the operator $Q$ acts from the domain $D$ into itself and is
uniformly contracting in the metric $\rho$ for some value of
$h>0$, provided that
\[
\overline{\omega}_1 = z_0;~~~ \overline{\omega}_2(x)
=-\overline{\omega}_1^2~~~{\rm for}~ |x| < 1;~~
\overline{\omega}_2(x) = \overline{\omega}_2(-x),~~{\it and}
\]
\[
\sup_{ \mathcal{X} \in \mathcal{C}_0} \sum_{x \in \mathcal{X}}
|\overline{\omega}_2(x)| \leq r \overline{\omega}_1^2.
\]
\end{lemma}
\proof Take $c = \frac{r +2}{3}$, $a_1 =\sqrt{\frac{2r}{r+1}}$,
$a_2 = 2$.
We shall need certain estimates on the values of $A(z,g)$ and
$B(z,g)$ for $(z,g) \in D$. Namely, there exist universal
constants $u_1,...,u_6$, such that for sufficiently small $z_0$ we
have
\begin{equation} \label{co1}
\sup_{(z,g) \in D}|A(z,g)| \leq u_1~.
\end{equation}
\begin{equation} \label{co2}
\sup_{(z,g) \in D}\sup_{ \mathcal{X}\in \mathcal{C}_0} \sum_{x \in
\mathcal{X} }|B(z,g)(x)| \leq u_2~.
\end{equation}
\begin{equation} \label{co3}
\sup_{(z_1,g), (z_2,g) \in D}|A(z_1,g)- A(z_2,g)| \leq u_3 |z_1 -
z_2| ~.
\end{equation}
\begin{equation} \label{co4}
\sup_{(z,g_1), (z,g_2) \in D}|A(z,g_1)- A(z,g_2)| \leq u_4
d(g_1,g_2)~.
\end{equation}
\begin{equation} \label{co5}
\sup_{(z_1,g), (z_2,g) \in D}\sup_{ \mathcal{X}\in \mathcal{C}_0}
\sum_{x \in \mathcal{X}}|B(z_1,g)(x)- B(z_2,g)(x)| \leq u_5 |z_1 -
z_2|~.
\end{equation}
\begin{equation} \label{co6}
\sup_{(z,g_1), (z,g_2) \in D}\sup_{ \mathcal{X}\in \mathcal{C}_0}
\sum_{x \in \mathcal{X}}|B(z,g_1)(x)- B(z,g_2)(x)| \leq u_6
d(g_1,g_2)~.
\end{equation}
These estimates follow from Lemma \ref{estu} below. For now,
assuming that they are true, we continue with the proof of Lemma
\ref{l2}. The fact that $Q D \subseteq D$ is guaranteed by the
inequalities
\begin{equation} \label{in1a}
z_0 + (a_2 z_0)^2 u_1 \leq a_2 z_0~,
\end{equation}
\begin{equation} \label{in2a}
z_0 - (a_2 z_0)^2 u_1 \geq a_1 z_0~,
\end{equation}
\begin{equation} \label{in3a}
\frac{r z_0^2}{(a_1 z_0)^2} + a_2 z_0 u_2 \leq c~.
\end{equation}
It is clear that (\ref{in1a})-(\ref{in3a}) hold for sufficiently
small $z_0$. Let us now demonstrate that  for some $h$ and for all
sufficiently small $z_0$ we have
\begin{equation} \label{contr}
\rho( Q(z_1,g_1), Q(z_2, g_2)) \leq \frac{1}{2} \rho( (z_1,g_1),
(z_2, g_2))~~~{\rm if}~~~  (z_1,g_1), (z_2,g_2) \in D.
\end{equation}
  First, taking (\ref{co1}), (\ref{co3}), and (\ref{co4}) into
account, we note that
\[
d_{z_0}(z_1^2 A(z_1,g_1), z_2^2 A(z_2,g_2)) \leq d_{z_0}(z_1^2
A(z_1,g_1), z_2^2 A(z_1,g_1)) +
\]
\[
d_{z_0}(z_2^2 A(z_1,g_1), z_2^2 A(z_2,g_1))+  d_{z_0}(z_2^2
A(z_2,g_1), z_2^2 A(z_2,g_2)) \leq
\]
\[
\frac{u_1 h |z_1^2 - z_2^2|}{z_0} + \frac{u_3 h (a_2 z_0)^2 |z_1 -
z_2|}{z_0} + \frac{u_4 h (a_2 z_0)^2 d(g_1,g_2)}{z_0}~.
\]
If  $h$ is fixed, the right hand side of this inequality can be
estimated from above, for all sufficiently small $z_0$, by
\[
\frac{1}{6}(d_{z_0}(z_1,z_2) + d(g_1,g_2)).
\]
Similarly,
\[
\sup_{ \mathcal{X}\in \mathcal{C}_0} \sum_{x \in \mathcal{X}}|z_1
B(z_1,g_1)(x) - z_2 B(z_2,g_2)(x)| \leq \sup_{ \mathcal{X}\in
\mathcal{C}_0} \sum_{x \in \mathcal{X}}|z_1 B(z_1,g_1)(x) - z_2
B(z_1,g_1)(x)| +
\]
\[
\sup_{ \mathcal{X}\in \mathcal{C}_0} \sum_{x \in \mathcal{X}}|z_2
B(z_1,g_1)(x) - z_2 B(z_2,g_1)(x)| + \sup_{ \mathcal{X}\in
\mathcal{C}_0} \sum_{x \in \mathcal{X}}|z_2 B(z_2,g_1)(x) - z_2
B(z_2,g_2)(x)| \leq
\]
\[
u_2|z_1 - z_2| + u_5 a_2 z_0 |z_1 - z_2| + u_6 a_2 z_0
d(g_1,g_2)~.
\]
Again, if $h$ is fixed, the right hand side of this inequality can
be estimated from above, for all sufficiently small $z_0$, by
\[
\frac{1}{6}(d_{z_0}(z_1,z_2) + d(g_1,g_2)).
\]
Finally, since $r <1$,
\[
\sup_{ \mathcal{X}\in \mathcal{C}_0} \sum_{x \in
\mathcal{X}}|\frac{\overline{\omega}_2(x)}{z_1^2} -
\frac{\overline{\omega}_2(x)}{z_2^2} | \leq r z_0^2
|\frac{1}{z_1^2} - \frac{1}{z_2^2}| \leq \frac{2 a_2 |z_1 -
z_2|}{a_1^4 z_0}~.
\]
 We can now take $h = \frac{12 a_2}{a_1^4}$, which implies that
the right hand side of the last inequality can be estimated from
above by $\frac{1}{6} d_{z_0}(z_1,z_2)$. We have thus demonstrated
the validity of~(\ref{contr}), which means that the operator $Q$
is uniformly contracting. This completes the proof of the lemma.
\qed
\section{Estimates on the Ursell Functions} \label{urs}
In this section we shall derive certain estimates on the Ursell
functions, which, in particular, will imply the inequalities
(\ref{co1})-(\ref{co6}).
\begin{lemma} \label{estu}
Suppose that the functions $g_1(x)$ and $ g_2(x)$ satisfy
(\ref{assum1})-(\ref{assum3}) with $a = c < 1$. Let
$\varphi^k=(\varphi^k_m(x_1,...,x_m))_{m \geq 0}$, $k = 1,2$ be
the corresponding Ursell functions. Then there exist constants
$q_1$ and $q_2$ such that
\[
\sup_{ \mathcal{Y}_1,..., \mathcal{Y}_n \in \mathcal{C}} \sum_{y_1
\in \mathcal{Y}_1... y_n \in \mathcal{Y}_n}
|\overline{\varphi}^k_{1+n}(y_1,...,y_n)| \leq n! q_1^{n+1}~,~~~k
= 1,2,
\]
\[
\sup_{ \mathcal{Y}_1,..., \mathcal{Y}_n \in \mathcal{C}} \sum_{y_1
\in \mathcal{Y}_1... y_n \in \mathcal{Y}_n}
|\overline{\varphi}^1_{1+n}(y_1,...,y_n)-
\overline{\varphi}^2_{1+n}(y_1,...,y_n) | \leq n! q_2^{n+1} ||g_1
-g_2||~.
\]
\end{lemma}
Note that the inequalities (\ref{co1})-(\ref{co6}) immediately
follow from this lemma, estimate (\ref{inte1}), and the
definitions of $A(z,g)$ and $B(z,g)(x)$.

Recall that in Section \ref{clust} we introduced the functions
$\widetilde{\varphi}_X(Y)$, which were closely related to the
Ursell functions. Given $g_1(x)$ and $ g_2(x)$ which satisfy
(\ref{assum1})-(\ref{assum3}) with $a = c < 1$, we now define
\[
r^k(m,n) = \sup_{(x_1,...,x_m)} \sup_{ \mathcal{Y}_1,...,
\mathcal{Y}_n \in \mathcal{C}} \sum_{y_1 \in \mathcal{Y}_1... y_n
\in \mathcal{Y}_n}
|\widetilde{\varphi}^k_{(x_1,...,x_m)}(y_1,...,y_n)|~,~~~k=1,2,
\]
\[
d(m,n) = \sup_{(x_1,...,x_m)} \sup_{ \mathcal{Y}_1,...,
\mathcal{Y}_n \in \mathcal{C}} \sum_{y_1 \in \mathcal{Y}_1... y_n
\in \mathcal{Y}_n}
|\widetilde{\varphi}^1_{(x_1,...,x_m)}(y_1,...,y_n)-
\widetilde{\varphi}^2_{(x_1,...,x_m)}(y_1,...,y_n)|~.
\]
We shall prove the following lemma.
\begin{lemma} \label{estu2}
Suppose that the functions $g_1(x)$ and $ g_2(x)$ satisfy
(\ref{assum1})-(\ref{assum3}) with $a=c < 1$.  Then there exist
constants $q_1$ and $q_2$ such that
\begin{equation} \label{er}
r^k(m,n) \leq n! q_1^{m+n}~,~~~k = 1,2,
\end{equation}
\begin{equation} \label{eq}
d(m,n) \leq n! q_2^{m+n} ||g_1 -g_2||~.
\end{equation}
\end{lemma}
Since we can express the Ursell functions in terms of
$\widetilde{\varphi}_X(Y)$ via (\ref{rel1}), Lemma \ref{estu2}
immediately implies Lemma \ref{estu}. It remains to prove Lemma
\ref{estu2}.\\ \\ {\it Proof of Lemma \ref{estu2}.} The estimate
(\ref{er}) can be obtained in the same way as (4.27) of \cite{A},
and thus we shall not prove it here. We proceed with the proof of
(\ref{eq}).

 In the definition of $d(m,n)$ we can take the
supremum over a restricted set of sequences $(x_1,...x_m)$, namely
those sequences, for which  $|x_j - x_i| \geq 1$ if $i \neq j$.
Indeed, if~${|x_j - x_i| < 1}$ for $i \neq j$, then
$\widetilde{\varphi}^1_{(x_1,...,x_m)}(y_1,...,y_n) =
\widetilde{\varphi}^2_{(x_1,...,x_m)}(y_1,...,y_n) = 0$, as
follows from the definition of $\widetilde{\varphi}_X(Y)$.

Let $f_k(x) =  e^{-\Phi_k(x)} = g_k (x) +1~$, $k =1,2$. For a
sequence $X = (x_1,...,x_m)$, let
\[
F_k^X = \prod_{i = 2}^m |f_k (x_i- x_1)|,~~~k  = 1,2, \] For a
sequence $Y = (y_1,...,y_s)$ and a point $x_1$, let
\[ G_k^{Y, x_1} = \prod_{j = 1}^s |g_k (y_j- x_1)|,~~~k  = 1,2. \]
 Note that
\[
F_k^{X} =
 \exp({  \sum_{i =2}^m \ln (g_k(x_i -x_1) +1) }) \leq
\]
\begin{equation} \label{fest}
  \leq \exp(  \sum_{i = 2}^m g_k(x_i -x_1) )\leq e^c ~~~~{\rm if}~|x_i - x_j |\geq 1 ~{\rm for}~ i \neq
  j.
\end{equation}
If $\mathcal{Y}_1,...,\mathcal{Y}_s \in \mathcal{C}$ then
\begin{equation}
\label{gest} \sum_{y_1 \in \mathcal{Y}_1... y_s \in \mathcal{Y}_s}
 G_k^{Y, x_1}
=  \prod_{j = 1}^s \sum_{ y \in \mathcal{Y}_j} |g_k (y- x_1)|
  \leq
(c_0+c)^s~,
\end{equation}
where $c_0$ is the largest number of points separated by unit
distance, which can fit inside a unit ball.

The proof of (\ref{eq}) will proceed via an induction on $m+n$.
Assume that $x_1,...,x_m$ are separated by unit distance. From the
recurrence relation (\ref{rec}) it follows that for $m \geq 1$
\[
 \sum_{y_1 \in \mathcal{Y}_1... y_n
\in \mathcal{Y}_n}
|\widetilde{\varphi}^1_{(x_1,...,x_m)}(y_1,...,y_n)-
\widetilde{\varphi}^2_{(x_1,...,x_m)}(y_1,...,y_n)| =
\]
\[
 \sum_{y_1 \in \mathcal{Y}_1... y_n
\in \mathcal{Y}_n} | \prod_{i = 2}^m f_1(x_i- x_1) \sum_{Z
\subseteq Y} \prod_{j =1}^{|Z|} g_1(z_j -x_1)
\widetilde{\varphi}^1_{Z \cup X \backslash x_1} (Y \backslash Z) -
\]
\[
 \prod_{i = 2}^m f_2(x_i- x_1) \sum_{Z \subseteq Y} \prod_{j = 1}^{|Z|}
  g_2(z_j -x_1) \widetilde{\varphi}^2_{Z \cup X \backslash x_1} (Y
\backslash Z)| \leq I_1 + I_2,
\]
where
\[
I_1 =   \sum_{y_1 \in \mathcal{Y}_1... y_n \in \mathcal{Y}_n}
\sum_{Z \subseteq Y} | \prod_{i = 2}^m f_1(x_i- x_1) \prod_{j =
1}^{|Z|} g_1(z_j -x_1) ( \widetilde{\varphi}^1_{Z \cup X
\backslash x_1} (Y \backslash Z) - \widetilde{\varphi}^2_{Z \cup X
\backslash x_1} (Y \backslash Z))|~,
\]
\[
I_2 =  \sum_{y_1 \in \mathcal{Y}_1... y_n \in \mathcal{Y}_n}
\sum_{Z \subseteq Y} |[ \prod_{i = 2}^m f_1(x_i- x_1)
\prod_{j=1}^{|Z|} g_1(z_j -x_1) -
\]
\[
  \prod_{i = 2}^m
f_2(x_i- x_1)  \prod_{j=1}^{|Z|} g_2(z_j -x_1)]
\widetilde{\varphi}^2_{Z \cup X \backslash x_1} (Y \backslash
Z)|~.
\]
Note that there are $\frac{n!}{s!(n-s)!}$ subsequences $Z$ of the
sequence $Y$, which are of length $s$. Rearranging the sum, so
that to take it first over all possible values of $s$, and then
over all possible subsequences of length $s$,  we see that
\[
I_1 \leq \sum_{s = 0}^n  \sum_{(k_1,...,k_s) \subseteq (1,...,n)}~
\sum_{z_1 \in \mathcal{Y}_{k_1}... z_s \in \mathcal{Y}_{k_s}}
 | \prod_{i = 2}^m f_1(x_i- x_1) \prod_{j =
1}^s g_1(z_j -x_1)| d(m+s -1, n-s) \leq
\]
\[
\sum_{s = 0}^n \frac{n!}{s!(n-s)!} \max_{(k_1,...,k_s) \subseteq
(1,...,n)}\sum_{z_1 \in \mathcal{Y}_{k_1}... z_s \in
\mathcal{Y}_{k_s}}
 | \prod_{i = 2}^m f_1(x_i- x_1) \prod_{j =
1}^s g_1(z_j -x_1)| d(m+s -1, n-s) \leq
\]
\[
 \sum_{s = 0}^n  \frac{n!}{s!(n-s)!} e^c(c_0+ c)^s  d(m+s -1, n-s)~.
\]
Similarly,
\[
I_2 \leq \sum_{s = 0}^n  \frac{n!}{s!(n-s)!} \max_{(k_1,...,k_s)
\subseteq (1,...,n)}\sum_{z_1 \in \mathcal{Y}_{k_1}... z_s \in
\mathcal{Y}_{k_s}}  | \prod_{i = 2}^m f_1(x_i- x_1) \prod_{j =
1}^s g_1(z_j -x_1) -
\]
\[
\prod_{i = 2}^m f_2(x_i- x_1) \prod_{j = 1}^s g_2(z_j -x_1) |
r(m+s-1, n-s)~.
\]

Then,
\[
 \sum_{z_1 \in
\mathcal{Y}_{k_1}... z_s \in \mathcal{Y}_{k_s}}  | \prod_{i = 2}^m
f_1(x_i- x_1) \prod_{j = 1}^s g_1(z_j -x_1) - \prod_{i = 2}^m
f_2(x_i- x_1) \prod_{j = 1}^s g_2(z_j -x_1) | \leq
\]
\[
 \sum_{z_1 \in
\mathcal{Y}_{k_1}... z_s \in \mathcal{Y}_{k_s}} [ |f_1(x_2-x_1) -
f_2(x_2 -x_1)|F_1^{(x_3,...,x_m)}G_1^{(z_1,...,z_s), x_1} +
\]
\[F_2^{(x_2)}|f_1(x_3-x_1) - f_2(x_3-x_1)| F_1^{(x_4,...,x_m)} G_1^{(z_1,...,z_s), x_1} +
...
\]
\[
...+F_2^{(x_2,...,x_{m-1})}|f_1(x_m - x_1) - f_2(x_m - x_1)|
G_1^{(z_1,...,z_s),x_1}+
\]
\[
F_2^{(x_2,...,x_m)}|g_1(z_1 - x_1) - g_2(z_1 - x_1)|
G_1^{(z_2,...,z_s),x_1} + ...
\]
\[
...+ F_2^{(x_2,...,x_m)} G_2^{(z_1,...,z_{s-1}), x_1} |g_1(z_s -
x_1) - g_2(z_s - x_1)| ].
\]
There are $m+s$ terms inside the square brackets. In addition to
(\ref{fest}) and (\ref{gest}), we use the fact that
\[
|f_1(x_i - x_1) - f_2(x_i - x_1)| \leq ||g_1 - g_2||~,~~~~2 \leq i
\leq m, \] \[
  \sum_{z_j \in
\mathcal{Y}_{k_j}} |g_1(z_j - x_1) - g_2(z_j - x_1)| \leq ||g_1 -
g_2||~,~~~~1 \leq j \leq s.
\]
Therefore, the entire sum can be estimated from above by
\[
(m+s) e^{2c} (c_0 +c)^s ||g_1 - g_2||.
\]
Therefore,
\[
I_2 \leq \sum_{s = 0}^n  \frac{n!}{s!(n-s)!}  (m+s) e^{2c} (c_0
+c)^s ||g_1 - g_2|| r(m+s-1, n-s) \leq
\]
\[
||g_1 - g_2|| (m+n) e^{2c} n! q_1^{m+n -1}  \sum_{s = 0}^n
\frac{(c_0+c)^s}{s!} \leq ||g_1 - g_2|| (m+n) e^{c_0+3c} n!
q_1^{m+n -1}~.
\]
Combining this with the estimate on $I_1$ we see that
\[
d(m,n) \leq  \sum_{s = 0}^n  \frac{n!}{s!(n-s)!} e^c(c_0+ c)^s
d(m+s -1, n-s) +
\]
\[
||g_1 - g_2|| (m+n) e^{c_0+3c} n! q_1^{m+n -1}~.
\]
Let us use induction on $m+n$ to prove that
\begin{equation}
\label{ind}  d(m,n) \leq n! q_2^{m+n} ||g_1 -g_2|| (m+n)
\end{equation}
for some value of $q_2$. The statement is  true when $m = 0$ since
$\widetilde{\varphi} = \mathbf{1}$ in this case. Assuming that the
induction hypothesis holds for all $m',n'$ with $m'+n' \leq m+n
-1$, we obtain for $m \geq 1$
\[
d(m,n) \leq  \sum_{s = 0}^n  \frac{n!}{s!(n-s)!} e^c(c_0+ c)^s
(n-s)! q_2^{m+n-1} ||g_1 -g_2||(m+n-1) +
\]
\[
||g_1 - g_2|| (m+n) e^{c_0+3c} n! q_1^{m+n -1} \leq
\]
\[
||g_1 - g_2|| (m+n) e^{c_0+3c} n!( q_1^{m+n -1} + q_2^{m+n -1}).
\]
The expression in the right hand side of this inequality is
estimated from above by the right hand side of (\ref{ind}) if $q_2
= 2 e^{c_0+3c} \max(1,q_1)$. Thus, (\ref{ind}) holds for all $m,n$
with this choice of $q_2$. Note that we can get rid of the factor
$(m+n)$ in the right hand side of (\ref{ind}) by taking a larger
value of $q_2$. This completes the proof of (\ref{eq}) and of
Lemma \ref{estu2}. \qed

\end{document}